\documentclass[Physsubmission, Phys]{SciPost}
\usepackage[T1]{fontenc}
\usepackage{lmodern}
\usepackage[utf8]{inputenc}
\usepackage{amsmath}
\usepackage{amssymb}
\usepackage{graphicx}
\usepackage{enumitem}
\usepackage[absolute]{textpos}
 \usepackage[affil-it]{authblk}
\usepackage{xcolor}
\usepackage{tabularx}
\usepackage{booktabs}
\usepackage{multirow}
\usepackage{listings}
\usepackage{bold-extra}

\hypersetup{
    colorlinks,
    linkcolor={red!50!black},
    citecolor={blue!50!black},
    urlcolor={blue!80!black}
}

\usepackage[bitstream-charter]{mathdesign}
\DeclareSymbolFont{usualmathcal}{OMS}{cmsy}{m}{n}
\DeclareSymbolFontAlphabet{\mathcal}{usualmathcal}

\newcommand*{\kira}{\code{Kira}}
\newcommand*{\kiraZO}{\code{Kira\;2.0}}

\newcommand*{\firefly}{\code{FireFly}}

\newcommand*{\fermat}{\code{Fermat}}

\newcommand{\code}[1]{\texttt{#1}}

\newcounter{notecount}

\renewcommand{\texttt}[1]{%
  \begingroup
  \ttfamily
  \begingroup\lccode`~=`|\lowercase{\endgroup\def~}{|\discretionary{}{}{}}%
  \begingroup\lccode`~=`_\lowercase{\endgroup\def~}{_\discretionary{}{}{}}%
  \catcode`|=\active\catcode`_=\active
  \scantokens{#1\noexpand}%
  \endgroup
}

\begin{document}

\begin{center}{\Large \textbf{
Developments since Kira 2.0\\
}}\end{center}

% TODO: write the author list here. Use initials + surname format.
% Separate subsequent authors by a comma, omit comma at the end of the list.
% Mark the corresponding author with a superscript *.
\begin{center}
Fabian Lange\textsuperscript{1,2},
Philipp Maierh\"ofer\textsuperscript{3} and
Johann Usovitsch\textsuperscript{4$\star$}
\end{center}

\begin{center}
{\bf 1} Institut f\"ur Theoretische Teilchenphysik, Karlsruhe Institute of Technology (KIT), Wolfgang-Gaede Straße 1, 76128 Karlsruhe, Germany
\\
{\bf 2} Institut f\"ur Astroteilchenphysik, Karlsruhe Institute of Technology (KIT), Hermann-von-Helmholtz-Platz 1, 76344 Eggenstein-Leopoldshafen, Germany
\\
{\bf 3} Physikalisches~Institut, Albert-Ludwigs-Universit\"at~Freiburg, 79104~Freiburg, Germany
\\
{\bf 4} Theoretical Physics Department, CERN, 1211 Geneva, Switzerland
\\
* johann.usovitsch@cern.com
\end{center}

\begin{center}
\today
\end{center}

\begin{center}
{\footnotesize CERN-TH-2021-178\quad{}TTP21-045\quad{} P3H-21-087}
\end{center}

% For convenience during refereeing (optional),
% you can turn on line numbers by uncommenting the next line:
%\linenumbers

\definecolor{palegray}{gray}{0.95}
\begin{center}
\colorbox{palegray}{
  \begin{tabular}{rr}
  \begin{minipage}{0.1\textwidth}
    \includegraphics[width=35mm]{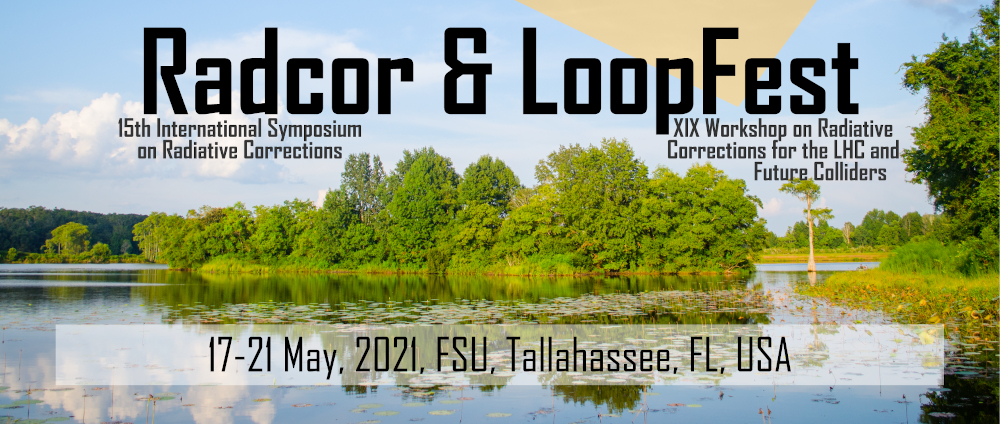}
  \end{minipage}
  &
  \begin{minipage}{0.85\textwidth}
    \begin{center}
    {\it 15th International Symposium on Radiative Corrections: \\Applications of Quantum Field Theory to Phenomenology,}\\
    {\it FSU, Tallahasse, FL, USA, 17-21 May 2021} \\
    \doi{10.21468/SciPostPhysProc.?}\\
    \end{center}
  \end{minipage}
\end{tabular}
}
\end{center}

\section*{Abstract}
{\bf
  Last year we released version \code{2.0} of the Feynman integral reduction program \code{Kira}.
  In this contribution we first report on changes and new features since then and, secondly, on new features for upcoming releases.
}

% \vspace{10pt}
% \noindent\rule{\textwidth}{1pt}
% \tableofcontents\thispagestyle{fancy}
% \noindent\rule{\textwidth}{1pt}
% \vspace{10pt}

% =================== %
\section{Introduction}
% =================== %

The reduction of Feynman integrals to a smaller set of master integrals is a crucial part of high-precision calculations in theoretical particle physics.
This can be achieved by using linear relations between integrals as provided by \emph{integration-by-parts} (IBP) identities~\cite{Tkachov:1981wb,Chetyrkin:1981qh} and \emph{Lorentz-invariance} (LI) identities~\cite{Gehrmann:1999as}.
To this end, \kira{} is an implementation of the Laporta algorithm~\cite{Laporta:2001dd} and was first published in Ref.~\cite{Maierhoefer:2017hyi}.
The algebraic expressions were solely handled by the computer algebra system \fermat{}~\cite{fermat}.
Last year, we released version \code{2.0} of \kira{}~\cite{Klappert:2020nbg}, which, as headline feature, additionally offers finite field interpolation and rational reconstruction techniques (see Refs.~\cite{vonManteuffel:2014ixa,Peraro:2016wsq} in the context of IBPs) provided by the library \code{FireFly}~\cite{Klappert:2019emp,Klappert:2020aqs}.
There are several other public implementations of the Laporta algorithm, see e.g.\ Refs.~\cite{Anastasiou:2004vj,vonManteuffel:2012np,Smirnov:2019qkx}.

In this contribution we highlight some of the new features already implemented since last year's release of version \code{2.0} in Sect.~\ref{sect:features_since_2.0} and then present some of the planned features for upcoming releases in Sect.~\ref{sect:features}.
We assume that the reader is familiar with \kira{} and refer to Refs.~\cite{Maierhoefer:2017hyi,Klappert:2020nbg} otherwise.

\section{New and changed features since Kira 2.0}
\label{sect:features_since_2.0}

In this section we highlight some of the new and changed features implemented in the versions \code{2.1} and \code{2.2} of \kira{}:
\begin{itemize}
  \item We introduced a version number for the database.
  Databases produced prior to \code{Kira\ 2.1} cannot be used for export jobs with \code{kira2form} and \code{kira2formfill} unless they are upgraded with the command line argument \code{--{}--force\_database\_format=fermat|firefly}.
  Three cases have to be distinguished:
  \begin{enumerate}
    \item If the database was produced with \fermat{}, use the command line argument\\ \code{--{}--force\_database\_format=fermat}.
    \item If the database was produced with \firefly{} without \code{insert\_prefactors}, use the command line argument \code{--{}--force\_database\_format=firefly}.
    \item If the database was produced with \firefly{} and \code{insert\_prefactors}, it cannot be upgraded.
    Instead, remove the database and rerun the job that produced it.
    If \code{firefly\_saves} still exists, most of the computational steps are skipped.
  \end{enumerate}
  Export jobs with \code{kira2math} and \code{kira2file} are not affected by this change.

  \item With the new \code{kira2formfill} job the results can now be exported as \code{fill} statements to fill \code{TableBase} objects in \code{FORM}~\cite{Vermaseren:2000nd}.

  \item The numerators of the rational functions exported with \code{kira2form} and \code{kira2formfill} are now enclosed by \code{num()}.

  \item The prefactors submitted through the option \code{insert\_prefactors} are now enclosed by \code{prefactor[]} when employing \code{kira2math} and by \code{prefactor()} when using \code{kira2form} and \code{kira2formfill}.

  \item We added the option \code{run\_initiate:\ masters} to stop the reduction after the master integrals have been identified without writing the system to disk.

  \item The weight bits for user-defined systems with indexed integrals \code{T[a,b,c,...]} are now automatically adjusted.
  Before, Kira crashed if the weights did not fit into the default representation.

  \item We changed the output format for the master integrals to make it more readable by both humans and other programs.
  This mainly affects the output to the console, \code{kira.log}, and \code{results/<TOPOLOGY>/masters(.final)}.

  \item  The command line option \code{--{}--parallel}/\code{--p} now accepts the arguments \code{physical} and \code{logical} to exploit all physical or logical cores, respectively.
  It is still possible to manually choose any number of threads.
  If the option is given to \kira{} without argument, it uses \code{physical} by default.
  Under \code{macOS}, \code{physical} refers to the number of logical cores.
\end{itemize}

Finally, we emphasize that we opened a wiki at
\begin{center}
  \url{https://gitlab.com/kira-pyred/kira/-/wikis/Home}
\end{center}
to serve as first contact point for users.
It already contains sections about best practice and troubleshooting.

% =================== %
\section{Features in upcoming releases}
% =================== %
\label{sect:features}

In this section we briefly present improved and new features for the upcoming releases of \kira{}.

% =================== %
\subsection{Faster export of the results}
% =================== %
\label{sect:export}

The export performance of the results to different formats, especially for \code{FORM}~\cite{Vermaseren:2000nd}, is improved significantly.
This is achieved by replacing the pipe to \code{Perl} by \code{C++} \code{std::regex}.
Tab.~\ref{tab:export} shows the improvements for the example \code{topo7}.
\begin{table}[ht]
  \begin{center}
    \caption{Export of results to different formats with and without reconstructing the mass set to $1$ for the example \code{topo7} with $r = 7$ and $s = 2$.
    The numbers are still preliminary.}
    \label{tab:export}
    {\renewcommand{\arraystretch}{1.3}
    \begin{tabular}{l|c c c}
      \toprule
      Mode & \code{Kira 2.2} & \code{Kira 2.X} & Improvement factor\\
      \midrule
      \code{--kira2math} & 0.22 s & 0.22 s & 1\\
      \midrule
      \code{--kira2form} & 12.21 s & 0.43 s & 28\\
      \midrule
      \begin{tabular}{@{}l@{}}\code{--kira2math} \\ \code{reconstruct\_mass:\ true}\end{tabular} &
      8.33 s & 2.41 s & 3.5 \\
      \midrule
      \begin{tabular}{@{}l@{}}\code{--kira2form} \\ \code{reconstruct\_mass:\ true}\end{tabular} &
      20.61 s & 2.61 s & 8\\
      \bottomrule
    \end{tabular}}
  \end{center}
\end{table}

\subsection{Reorder propagators to increase the performance}
\label{sect:reorder}

The definition of the topology can have an impact on the size of the system generated by \kira{} and on the performance of the reduction.
Not only redefinitions of the propagators can change the performance, even a simple reordering can drastically impact it.
In Ref.~\cite{Maierhofer:2018gpa} we described rough guidelines for the topology definition, which we repeat in the following for clarity:
\begin{enumerate}
  \item Always define \code{top\_level\_sectors} for the topologies.
  This ensures that sectors are only mapped on subsectors of the defined top-level sectors.
  If you are trying to find relations by reducing sectors which are higher than those in the physical problem, the higher sectors must be included in the top-level sectors or the \code{magic\_relations} have to be enabled in order to symmetrize them.
  \item If a topology has only one top-level sector, order the propagators such that the sector number is $2^n - 1$ (if the sector has $n$ lines), i.e.\ the propagators
at the end of the list appear only as irreducible scalar products.
  \item Define the propagators in such a way that propagators have short linear combinations of loop momenta and the shortest linear combinations appear earlier in the list of propagators.
  \item Analogously, keep linear combinations of external momenta short with ascending length.
  \item Massless propagators should appear earlier in the list.
\end{enumerate}
These guidelines are based on experience and will not always produce the best definition.
However, they should serve as a good starting point.
If the topology definition is crucial to complete the reduction, one should run smaller jobs first to experimentally find the preferred definition.

\begin{figure}[ht]
  \centering
  \includegraphics[trim = 36 538 36 36, scale = 0.5]{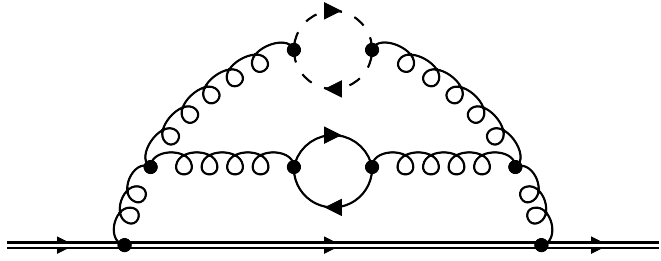}
  \caption{Example topology from Ref.~\cite{Fael:2021kyg}.
  The double line represents quarks with mass $m_1$, the single line quarks with mass $m_2$, and the dashed line massless quarks.
  Produced with \code{FeynGame}~\cite{Harlander:2020cyh}.}
  \label{fig:d4L408}
\end{figure}

In Ref.~\cite{Fael:2021kyg} one of the authors of this contribution encountered some extreme examples in the calculation of four-loop on-shell integrals for the relation of a heavy quark defined in the $\overline{\mathrm{MS}}$ and the on-shell scheme allowing for a second non-zero quark mass in the loops.
We show one example topology in Fig.~\ref{fig:d4L408}.
The setup used in Ref.~\cite{Fael:2021kyg} automatically defines its propagators as
\begin{equation}
  \begin{gathered}
    P_1 = p_1^2 - m_1^2, \quad
    P_2 = (p_1 - q)^2, \quad
    P_3 = (p_2 + p_3)^2, \quad
    P_4 = (p_1 - p_3 + p_4 - q)^2, \\
    P_5 = (p_3 - p_4)^2, \quad
    P_6 = (p_3 + q)^2, \quad
    P_7 = (p_2 + p_4)^2, \quad
    P_8 = p_2^2 - m_2^2, \\
    P_9 = (-p_1 + p_2 + p_3 - p_4 + q)^2 - m_2^2, \quad
    P_{10} = p_3^2, \quad
    P_{11} = p_4^2, \quad
    P_{12} = (p_1 + p_3)^2, \\
    P_{13} = (p_1 + p_4)^2, \quad
    P_{14} = (p_1 + p_2)^2,
  \end{gathered}
\end{equation}
where $p_i$ are the loop momenta, $q$ is the external momentum, and $m_1$, $m_2$ are the two quark masses.
$P_3$, $P_6$, $P_7$, $P_{12}$, $P_{13}$, and $P_{14}$ are auxiliary propagators which only appear in the numerators.
With this definition, \code{Kira} generates $10\,204\,834$ linearly independent equations with $436\,973\,786$ terms in total to reduce the selected integrals and one reduction over a finite field with \code{pyRed} takes about $12\,771$\,s.
Following the guidelines above and reordering the propagators to
\begin{equation}
  \begin{gathered}
    P_{1} = p_3^2, \quad
    P_{2} = p_4^2, \quad
    P_3 = p_1^2 - m_1^2, \quad
    P_4 = p_2^2 - m_2^2, \quad
    P_5 = (p_1 - q)^2, \quad
    P_6 = (p_3 - p_4)^2, \\
    P_7 = (p_1 - p_3 + p_4 - q)^2, \quad
    P_8 = (-p_1 + p_2 + p_3 - p_4 + q)^2 - m_2^2, \quad
    P_{9} = (p_3 + q)^2, \\
    P_{10} = (p_2 + p_4)^2, \quad
    P_{11} = (p_2 + p_3)^2, \quad
    P_{12} = (p_1 + p_3)^2, \quad
    P_{13} = (p_1 + p_4)^2, \\
    P_{14} = (p_1 + p_2)^2
  \end{gathered}
\end{equation}
only $3\,752\,490$ equations with $57\,696\,557$ terms are generated, corresponding to a reduction by $63$\,\% and $87$\,\%, respectively.
The runtime of a reduction with \code{pyRed} reduces to $43$\,s, i.e. the performance increases by a factor of $297$.
We want to stress that this is certainly an extreme example and the performance gains can even be negligible for other examples.

For the convenience of the users of \kira{}, we plan to implement the option to reorder the propagators internally, i.e.\ to allow the internal order to differ from the definition in \code{integralfamilies.yaml}.
This allows for reaping the potential performance gains without keeping track of the order when communicating with other programs.
We plan to implement two versions:
first, allow the users to set the ordering manually and, secondly, automatically reorder the propagators according to the guidelines above.

\subsection{Generate a system of equations for later reduction as user-defined system}
\label{sect:generate_input}

With the option \code{run\_initiate:\ input}, the generated system of linearly independent equations is stored in the \kira{} readable \code{.kira} format for further processing.
In contrast to the existing option \code{generate\_input}, which was introduced to reduce the memory requirements by considering only a subset of the sectors at the same time, with the option \code{run\_initiate:\ input} the user is able to select equations to reduce only requested integrals.
One idea is to use these files \code{.kira} in combination with additionally generated \code{.kira} files containing extra relations provided by the user.
The additional relations can be extra symmetry relations, a system of differential equations, or amplitudes.

\subsection{Set variables to rational numbers and/or choose a prime field}
\label{sect:set_value}

The complexity of reductions can be greatly reduced by setting some or all of the variables to rational numbers.
While the result is no longer general, it might be helpful to first reduce the simpler problem to get a feeling for the reduction and the structure of the result.
We plan to allow the users to set the variables to rational numbers at different stages of \kira{} so that, for example, the system can be generated in full generality and then be solved with the variables replaced by rational numbers.

Additionally, we want to allow the users to choose a prime and solve the system one or several times over that prime field.
By storing these results to disk, they could then be processed by other programs if the users are only interested in the solutions over specific prime fields but not the algebraic result.

\section{Conclusions}
\label{sect:conclusions}

In this contribution we presented the new and changed features already implemented since the release of \kiraZO{} and some of the features planned for upcoming releases.
We hope that these, together with the ever-present bug fixes, will improve the experience of users of \kira{}.

% ====================== %
\section*{Acknowledgments}
% ====================== %

\paragraph{Funding information}

The research of F.L.\ was supported by the \textit{Deutsche Forschungsgemeinschaft} (DFG, German Research Foundation) through the Collaborative Research Centre \href{http://p3h.particle.kit.edu/start}{TRR 257} funded through grant \href{http://gepris.dfg.de/gepris/projekt/396021762?language=en}{396021762}.

\bibliographystyle{SciPost_bibstyle}
\bibliography{literature}

\begin{thebibliography}{10}
\providecommand{\url}[1]{\texttt{#1}}
\providecommand{\urlprefix}{URL }
\expandafter\ifx\csname urlstyle\endcsname\relax
  \providecommand{\doi}[1]{doi:\discretionary{}{}{}#1}\else
  \providecommand{\doi}{doi:\discretionary{}{}{}\begingroup
  \urlstyle{rm}\Url}\fi
\providecommand{\eprint}[2][]{\url{#2}}

\bibitem{Tkachov:1981wb}
F.~V. Tkachov,
\newblock \emph{{A theorem on analytical calculability of 4-loop
  renormalization group functions}},
\newblock Phys. Lett. B \textbf{100}, 65 (1981),
\newblock \doi{10.1016/0370-2693(81)90288-4}.

\bibitem{Chetyrkin:1981qh}
K.~G. Chetyrkin and F.~V. Tkachov,
\newblock \emph{{Integration by parts: The algorithm to calculate
  $\beta$-functions in 4 loops}},
\newblock Nucl. Phys. \textbf{B192}, 159 (1981),
\newblock \doi{10.1016/0550-3213(81)90199-1}.

\bibitem{Gehrmann:1999as}
T.~Gehrmann and E.~Remiddi,
\newblock \emph{{Differential equations for two-loop four-point functions}},
\newblock Nucl. Phys. \textbf{B580}, 485 (2000),
\newblock \doi{10.1016/S0550-3213(00)00223-6},
\newblock \eprint{hep-ph/9912329}.

\bibitem{Laporta:2001dd}
S.~Laporta,
\newblock \emph{{High-precision calculation of multiloop Feynman integrals by
  difference equations}},
\newblock Int.J.Mod.Phys. \textbf{A15}, 5087 (2000),
\newblock \doi{10.1016/S0217-751X(00)00215-7},
\newblock \eprint{hep-ph/0102033}.

\bibitem{Maierhoefer:2017hyi}
P.~Maierhöfer, J.~Usovitsch and P.~Uwer,
\newblock \emph{{Kira—A Feynman integral reduction program}},
\newblock Comput. Phys. Commun. \textbf{230}, 99 (2018),
\newblock \doi{10.1016/j.cpc.2018.04.012},
\newblock \eprint{1705.05610}.

\bibitem{fermat}
R.~H. Lewis,
\newblock \emph{{Computer Algebra System Fermat}},
\newblock \urlprefix\url{https://home.bway.net/lewis}.

\bibitem{Klappert:2020nbg}
J.~Klappert, F.~Lange, P.~Maierh\"ofer and J.~Usovitsch,
\newblock \emph{{Integral reduction with Kira 2.0 and finite field methods}},
\newblock Comput. Phys. Commun. \textbf{266}, 108024 (2021),
\newblock \doi{10.1016/j.cpc.2021.108024},
\newblock \eprint{2008.06494}.

\bibitem{vonManteuffel:2014ixa}
A.~von Manteuffel and R.~M. Schabinger,
\newblock \emph{{A novel approach to integration by parts reduction}},
\newblock Phys. Lett. \textbf{B744}, 101 (2015),
\newblock \doi{10.1016/j.physletb.2015.03.029},
\newblock \eprint{1406.4513}.

\bibitem{Peraro:2016wsq}
T.~Peraro,
\newblock \emph{{Scattering amplitudes over finite fields and multivariate
  functional reconstruction}},
\newblock JHEP \textbf{12}, 030 (2016),
\newblock \doi{10.1007/JHEP12(2016)030},
\newblock \eprint{1608.01902}.

\bibitem{Klappert:2019emp}
J.~Klappert and F.~Lange,
\newblock \emph{{Reconstructing rational functions with FireFly}},
\newblock Comput. Phys. Commun. \textbf{247}, 106951 (2020),
\newblock \doi{10.1016/j.cpc.2019.106951},
\newblock \eprint{1904.00009}.

\bibitem{Klappert:2020aqs}
J.~Klappert, S.~Y. Klein and F.~Lange,
\newblock \emph{{Interpolation of dense and sparse rational functions and other
  improvements in FireFly}},
\newblock Comput. Phys. Commun. \textbf{264}, 107968 (2021),
\newblock \doi{10.1016/j.cpc.2021.107968},
\newblock \eprint{2004.01463}.

\bibitem{Anastasiou:2004vj}
C.~Anastasiou and A.~Lazopoulos,
\newblock \emph{{Automatic integral reduction for higher order perturbative
  calculations}},
\newblock JHEP \textbf{07}, 046 (2004),
\newblock \doi{10.1088/1126-6708/2004/07/046},
\newblock \eprint{hep-ph/0404258}.

\bibitem{vonManteuffel:2012np}
A.~von Manteuffel and C.~Studerus,
\newblock \emph{{Reduze 2 -- Distributed Feynman Integral Reduction}}  (2012),
\newblock \eprint{1201.4330}.

\bibitem{Smirnov:2019qkx}
A.~V. Smirnov and F.~S. Chukharev,
\newblock \emph{{FIRE6: Feynman Integral REduction with modular arithmetic}},
\newblock Comput. Phys. Commun. \textbf{247}, 106877 (2020),
\newblock \doi{10.1016/j.cpc.2019.106877},
\newblock \eprint{1901.07808}.

\bibitem{Vermaseren:2000nd}
J.~A.~M. Vermaseren,
\newblock \emph{{New features of FORM}}  (2000),
\newblock \eprint{math-ph/0010025}.

\bibitem{Maierhofer:2018gpa}
P.~Maierhöfer and J.~Usovitsch,
\newblock \emph{{Kira 1.2 Release Notes}}  (2018),
\newblock \eprint{1812.01491}.

\bibitem{Fael:2021kyg}
M.~Fael, F.~Lange, K.~Sch\"onwald and M.~Steinhauser,
\newblock \emph{{A semi-analytic method to compute Feynman integrals applied to
  four-loop corrections to the $\overline{\rm MS}$-pole quark mass relation}},
\newblock JHEP \textbf{09}, 152 (2021),
\newblock \doi{10.1007/JHEP09(2021)152},
\newblock \eprint{2106.05296}.

\bibitem{Harlander:2020cyh}
R.~V. Harlander, S.~Y. Klein and M.~Lipp,
\newblock \emph{{FeynGame}},
\newblock Comput. Phys. Commun. \textbf{256}, 107465 (2020),
\newblock \doi{10.1016/j.cpc.2020.107465},
\newblock \eprint{2003.00896}.

\end{thebibliography}

\nolinenumbers

\end{document}